\documentclass[preprint,showpacs,preprintnumbers,amsmath,amssymb]{revtex4}

\usepackage{graphicx}
\usepackage{dcolumn}
\usepackage{amsmath}
\usepackage{amssymb}

\def\vk{{\bf k}}

\def\bra{\langle}
\def\ket{\rangle}

\newcommand{\fig}[1]{Fig.~\ref{#1}}
\newcommand{\be}{\begin{equation}}
\newcommand{\ee}{\end{equation}}
\newcommand{\bea}{\begin{eqnarray}}
\newcommand{\no}{\nonumber}
\newcommand{\eea}{\end{eqnarray}}
\newcommand{\bean}{\begin{eqnarray*}}
\newcommand{\eean}{\end{eqnarray*}}
\newcommand{\bfi}{\begin{figure}}
\newcommand{\efi}{\end{figure}}
\newcommand{\bc}{\begin{center}}
\newcommand{\ec}{\end{center}}
\newcommand{\ba}{\begin{array}}
\newcommand{\ea}{\end{array}}

\begin{document}


\title{Spontaneous Fermi surface symmetry breaking in bilayered systems}

\author{Hiroyuki Yamase} 
\affiliation{National Institute for Materials Science, 1-2-1, Sengan, Tsukuba 305-0047, Japan} 
\affiliation{Max-Planck-Institute for Solid State Research, 
Heisenbergstrasse 1, D-70569 Stuttgart, Germany}



\date{June 3, 2009}

\begin{abstract} 
We perform a comprehensive numerical study of $d$-wave Fermi surface deformations ($d$FSD) 
on a square lattice, the so-called $d$-wave Pomeranchuk instability, including bilayer coupling. 
Since the order parameter corresponding to the $d$FSD has Ising symmetry, 
there are  two stacking patterns between the layeres, $(+,+)$ and $(+,-)$. 
This additional degree of freedom gives rise to a rich variety of phase diagrams. 
The phase diagrams are classified by means of the energy scale $\Lambda_{z}$, which 
is defined as the bilayer splitting at the saddle points of  the in-plane band dispersion. 
As long as $\Lambda_{z}\ne 0$, a major stacking pattern is usually $(+,-)$, 
and $(+,+)$ stacking is stabilized  as a dominant pattern only when 
the temperature scale of the $d$FSD instability becomes much smaller than $\Lambda_z$. 
For $\Lambda_{z}=0$, 
the phase diagram depends on the precise form of the bilayer dispersion. 
We also analyze 
the effect of a magnetic field on the bilayer model 
in connection with a possible $d$FSD instability in the bilyared ruthenate Sr$_{3}$Ru$_{2}$O$_{7}$. 
\end{abstract}

\maketitle

\section{Introduction}  
\begin{figure}
\centerline{\includegraphics[width=0.45\textwidth]{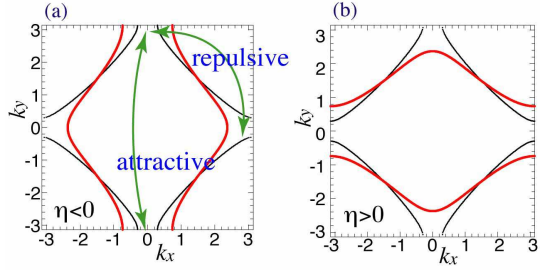}}
\caption{(Color online) $d$-wave Fermi surface deformations ($d$FSD). 
Forward scattering processes of quasiparticles around $(\pi,0)$ and $(0,\pi)$ 
drive symmetry breaking 
of the Fermi surface (red lines). This symmetry breaking is 
characterized by the order parameter $\eta$, 
which is negative and positive in (a) and  (b), respectively.  
}
\label{dFSD}
\end{figure}

While a wide variety of shapes of the Fermi surface are realized in metals,  
the Fermi surface usually respects 
the point-group symmetry of the underlying lattice structure. 
However, it was found that Fermi surface symmetry 
can be broken spontaneously due to electron-electron correlations in the two-dimensional 
$t$-$J$,\cite{yamase00a,yamase00b}  Hubbard,\cite{metzner00} and 
extended Hubbard\cite{valenzuela01} models. 
This instability is driven by forward scattering processes of quasi-particles. 
Standard model interaction leading to such symmetry breaking is given by 
\be
\sum_{\vk, \vk'} f_{\vk \vk'} n_{\vk} n_{\vk'} \, .
\label{fkk}
\ee
Here 
\be
 f_{\vk \vk'}= -g d_{\vk} d_{\vk'}
 \label{fkk2}
 \ee
is the forward scattering interaction with $d$-wave symmetry 
$d_{\vk}=\cos k_x - \cos k_y$ and the coupling constant $g>0$; 
$n_{\vk}$ is the electron density operator. 
The interaction (\ref{fkk}) gives rise to attraction between 
quasi-particles around $(0,\pi)$ and those around $(0,-\pi)$, 
and repulsion between $(0,\pi)$ and $(\pi,0)$. 
As a result, symmetry of the Fermi surface may be broken spontaneously at low temperature 
as shown by the red lines in \fig{dFSD}. 
These $d$-wave Fermi surface deformations ($d$FSD) 
break orientational symmetry of a square lattice and 
are often called 
a $d$-wave Pomeranchuk instability or an electronic nematic transition. 
While these three phrases are currently used in the same meaning, 
it may be worth mentioning the conceptional difference. 
The Pomeranchuk instability indicates breaking of Pomeranchuk's stability criterion for 
isotropic Fermi liquids.\cite{pomeranchuk58} 
However the $d$FSD instability can occur also for strongly correlated electron systems 
such as those described by the $t$-$J$ model.\cite{yamase00a,yamase00b,miyanaga06,edegger06} 
Moreover, the $d$FSD instability can be realized without breaking 
Pomeranchuk's criterion, because the transition is typically of first order 
at low temperature.\cite{kee0304,yamase05} 
The concept of the electronic nematic state was originally introduced to describe 
melting of possible charge stripes in cuprate superconductors.\cite{kivelson98} 
Hence the electronic nematic state often implies underlying charge-stripe order.\cite{kivelson03} 
However, the $d$FSD is driven by forward scattering interactions, 
not by the underlying charge stripes which necessarily generate a finite momentum transfer. 
The $d$FSD instability provides a different route to 
the electronic nematic state without assuming the underlying charge-stripe order.

The minimal model describing the $d$FSD instability consists of the forward scattering 
interaction (\ref{fkk}) and a kinetic term of electrons. 
This model, which we refer to as the f-model, 
was extensively investigated in Refs.~\cite{kee0304,yamase05,lamas08}. 
While the interaction considered in Ref.~\onlinecite{kee0304} is expressed in terms of 
quadrupole density,  
it becomes the same as our interaction (\ref{fkk}) after a mean-field 
calculation in Ref.~\onlinecite{kee0304}.    
The $d$FSD instability occurs around the van Hove filling 
with a dome-shaped transition line in a plane of the chemical potential and temperature. 
The transition is of second order around the center of the dome and changes 
to a first order at the edges of the dome; the end points of the second order line 
are tricritical points. 
In the weak coupling limit, the phase diagram is characterized by a single energy 
scale, yielding various universal ratios.\cite{yamase05}

The double-layered strontium ruthenate Sr$_{3}$Ru$_{2}$O$_{7}$ 
is a material possibly exhibiting the $d$FSD instability,\cite{grigera04,borzi07}  
which was supported by theoretical studies.\cite{kee05,doh07,yamase07b,yamase07c,ho08}  
The idea of the $d$FSD was also invoked in the context of high-$T_{c}$ cuprates\cite{yamase00a,yamase00b,yamase06}  
to understand the strong $xy$ anisotropy of magnetic excitation spectra 
in the underdoped and optimally doped YBa$_{2}$Cu$_{3}$O$_{y}$ 
with $y=6.6$ and $6.85$.\cite{hinkov04,hinkov07} 
In the  more underdoped material YBa$_{2}$Cu$_{3}$O$_{6.45}$, 
much stronger anisotropy was observed\cite{hinkov08}  and 
two scenarios were proposed: 
(i) a qunatum phase transition to the $d$FSD deeply inside the 
$d$-wave superconducting state\cite{kim08,huh08} and 
(ii) strong suppression of singlet pairing, which concomitantly enhances the 
$d$FSD order, since the $d$FSD is order competing with singlet pairing.\cite{yamase09} 
The two-dimensional electron gas is also known to show strong anisotropy 
of resistivity at low temperature in half-filled higher Landau levels.\cite{lilly99, du99} 
The orientation of the anisotropy always appears along the crystallographic direction.\cite{lilly99a} 
Theoretically the observed anisotropy was interpreted as nematic order in 
continuum models.\cite{fradkin99,fradkin00,wexler01}  

So far, no other materials exhibiting the $d$FSD instability are known. 
However, the $d$FSD is a generic tendency in correlated electron systems. 
It was found not only in the $t$-$J$\cite{yamase00a,yamase00b} and Hubbard\cite{metzner00,valenzuela01} models 
but also in more general models with central particle-particle interactions.\cite{quintanilla08} 
The $d$FSD can also occur in a three-dimensional system.\cite{quintanilla06} 
Therefore the $d$FSD is an interesting possibility for various materials, 
except if other instabilities prevail over it.

\begin{figure}
\centerline{\includegraphics[width=0.4\textwidth]{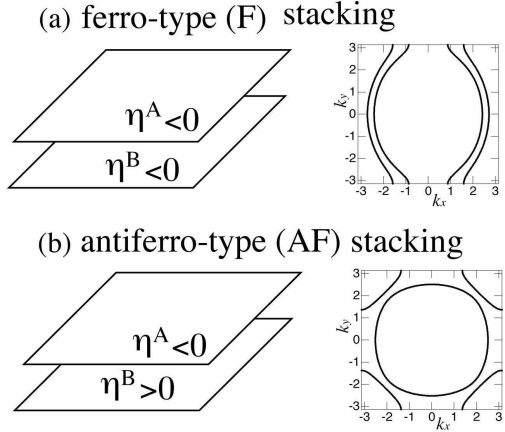}}
\caption{Stacking patterns of the $d$FSD in a bilayer system 
and the Fermi surfaces for $\epsilon_{\vk}^{z}=-t_{z}$: 
(a) ferro-type (F) stacking ($\eta^{A}\eta^{B}>0$) and 
(b) antiferro-type (AF) stacking ($\eta^{A}\eta^{B}<0$). 
A large $t_{z} (=0.3)$ is employed so that the bilayer splitting becomes apparent. 
}
\label{stacking}
\end{figure}

In layered materials, weak interlayer coupling is present. 
Since  the order parameter of the $d$FSD is characterized by Ising symmetry (see \fig{dFSD}), 
there are two stacking patterns $(+,+)$ and $(+,-)$,  as shown in \fig{stacking}; 
we call the former {\it ferro-type} (F) stacking and the latter {\it antiferro-type} (AF). 
In the latter case, macroscopic anisotropy does not appear [\fig{stacking}(b)], 
leading to self-masking of the underlying $d$FSD instability. 
In the framework of the Landau expansion of the free energy, 
it was found that AF stacking is usually favored 
as long as the $c$ axis dispersion at the saddle points of the 
in-plane band dispersion is finite.\cite{yamase09a}. 
That is, the $d$FSD turns out to provide spontaneous symmetry breaking 
which is usually self-masked in layered systems. The study Ref.~\onlinecite{yamase09a} 
suggests a possibility that the $d$FSD  is hidden in various materials. 

In this paper, we perform a comprehensive numerical study of the $d$FSD instability 
in the bilayer f-model. 
We show that the inclusion of bilayer coupling in the f-model yields a rich variety of 
phase diagrams upon tuning interaction strength, bilayer dispersions, and hopping integrals. 
The important quantity is the energy scale $\Lambda_{z}$, 
the bilayer splitting at the saddle points of the in-plane band dispersion. 
For $\Lambda_{z}\ne 0$, a major stacking pattern is usually AF, and 
F stacking is stabilized  as a dominant pattern 
only when the temperature scale of the $d$FSD 
becomes much smaller than $\Lambda_{z}$. 
For $\Lambda_{z}=0$, the phase diagram 
depends strongly on the form of the bilayer dispersion, leading to a variety of phase diagrams. 
While the saddle points are frequently located at $(\pi,0)$ and $(0,\pi)$ 
in a square lattice system, they may shift to other $\vk$ points in the presence of 
long-range hopping integrals. Even in this case, we demonstrate 
that our conclusion holds. 
Considering that the $d$FSD instability is likely to occur upon applying a magnetic field in 
the bilayered ruthenate Sr$_{3}$Ru$_{2}$O$_{7}$,\cite{grigera04,borzi07}  
we also calculate the phase diagram including the field in the bilayer f-model 
and choosing parameters appropriate to Sr$_{3}$Ru$_{2}$O$_{7}$. 
F stacking is stabilized around the van Hove energy of the bonding band, 
but the $d$FSD around that of the antibonding band is found to be strongly suppressed 
by the field. 

The paper is structured as follows. 
In Sec.II we introduce the bilayer f-model. Phase diagrams are 
presented in Sec.III for various choices of coupling strength, bilayer dispersions, and long-range 
hopping integrals. 
We also show a phase diagram in the presence of a magnetic field, imitating the experimental 
situation in Sr$_{3}$Ru$_{2}$O$_{7}$. 
The conclusions follow in Sec.IV. 
The present work is complementary to Ref.~\onlinecite{yamase09a} and 
elucidates possible phase diagrams of the $d$FSD instability in the bilayer model. 
We hope it will serve  as a sound foundation to explore the $d$FSD instability in bilayered systems.

\section{Model and formalism} 
We focus on the minimal bilayer model exhibiting the $d$FSD instability and analyze 
the following Hamiltonian, 
\bea
&&H = \sum_{\substack{\vk,\, \sigma \\ i=A,B}} 
(\epsilon_{\vk} - \mu) 
c^{i\,\dagger}_{\vk\sigma} c^{i}_{\vk\sigma} + 
 \frac{1}{2N} \sum_{\substack{\vk,\, \vk' \\ i=A,B}} 
f_{\vk\vk'} \, n^{i}_{\vk} n^{i}_{\vk'}  \no \\
&&\qquad\qquad\qquad +\sum_{\vk,\,\sigma} \epsilon_{\vk}^{z} (
c^{A\,\dagger}_{\vk \sigma} c^{B}_{\vk \sigma} + 
c^{B\,\dagger}_{\vk \sigma} c^{A}_{\vk \sigma}
)\,, \label{model}
\eea 
where $c^{i\,\dagger}_{\vk\sigma} (c^{i}_{\vk\sigma})$ 
creates (annihilates) an electron with momentum $\vk$ and 
spin $\sigma$ in the $i=A$ and $B$ planes; 
$n^{i}_{\vk} = \sum_{\sigma}c^{i\,\dagger}_{\vk\sigma} 
c^{i}_{\vk\sigma}$ is the number operator; 
$N$ is the total number of sites on the $i$ plane and 
$\mu$ denotes the chemical potential. 
We consider hopping amplitudes up to third nearest neighbors, i.e., $t$, $t'$, and $t''$, 
on the square lattice. The in-plane band dispersion $\epsilon_{\vk}$ is thus given by 
\be
 \epsilon_{\vk} = -2 t (\cos k_{x}+\cos k_{y}) 
 - 4t'\cos k_{x} \cos k_{y} 
 -2 t''(\cos 2k_{x}+\cos 2k_{y})\; . 
\ee
The forward scattering interaction $f_{\vk \vk'}$ 
drives the $d$FSD instability as shown in \fig{dFSD}. 
This interaction mimics the effective interaction obtained in 
the $t$-$J$\cite{yamase00a,yamase00b} and Hubbard\cite{metzner00,valenzuela01} models.
The last term in Hamiltonian (\ref{model}) is the hybridization between 
A and B planes. 
We consider four types of bilayer dispersions, 
$\epsilon_{\vk}^{z}=
-\frac{t_{z}}{4} (\cos k_x - \cos k_y)^{2}$, 
$-2t_{z} (\cos k_x + \cos k_y)$, 
$-4t_{z} \cos\frac{k_x}{2} \cos\frac{k_y}{2}$, 
and $-t_{z}$; 
The first one is the dispersion suggested for bilayer cuprates such as 
YBa$_{2}$Cu$_{3}$O$_{y}$;\cite{andersen95} the second is a dispersion taking account of 
next nearest-neighbor hopping between layers; the third is an expected dispersion 
in a system where adjacent layers are shifted by $(\frac{1}{2}, \frac{1}{2})$; 
the forth is the simplest one.

Hamiltonian (\ref{model}) is analyzed in the Hartree approximation, 
which becomes the exact analysis of our model in the thermodynamic limit. 
We obtain the mean field 
\be
\eta^{A (B)}=-\frac{g}{N} \sum_{\vk} d_{\vk} \bra n_{\vk}^{A (B)} \ket \,, 
\ee
which is nonzero only if the electronic state loses fourfold 
symmetry of the square lattice and is thus the order parameter of the 
$d$FSD in the $A (B)$ plane. 
The FS is elongated along the $k_{x}$ and $k_{y}$ directions 
for $\eta^{A(B)}>0$ and $\eta^{A(B)}<0$, respectively, as shown in \fig{dFSD}, 
i.e., the order parameter has Ising symmetry. F (AF) stacking 
is thus defined by $\eta^{A} \eta^{B}>0 (<0)$ (see \fig{stacking}).  
The mean-field Hamiltonian reads 
\be
\hspace{-0mm}H_{\rm MF}=\sum_{\vk,\, \sigma}
\left(
      c_{\vk \sigma}^{A\,\dagger}\;\;  c_{\vk \sigma}^{B\,\dagger} 
\right) 
\left( \begin{array}{cc} 
   \xi_{\vk}^{A} & \epsilon_{\vk}^{z} \\
\epsilon_{\vk}^{z} & \xi_{\vk}^{B} 
          \end{array}\right)
\left( \begin{array}{c}
 c_{\vk \sigma}^{A} \\
 c_{\vk \sigma}^{B} 
\end{array}\right)
+ \frac{N}{2g}[(\eta^{A})^{2}+(\eta^{B})^{2}] 
\, \label{MFH}
\ee
where $\xi^{A(B)}_{\vk} = \epsilon_{\vk}+\eta^{A(B)}d_{\vk}-\mu$. 
We determine the mean fields self-consistently under 
the constraint that each plane has the same electron density. 
A solution with $|\eta^{A}| \neq |\eta^{B}|$ is in principle allowed and 
induces spontaneous charge imbalance between 
the planes.\cite{miscchargeimbalance} 
However, such a solution costs energy by producing an electric field between 
the planes. 
The bilayer coupling is generally expected to be weak in layered materials and thus 
we fix $t_z/t=0.1$.

\section{Results} 
In a square lattice system, the saddle points are located in $(\pi,0)$ and $(0,\pi)$ 
for $|t'/t|<0.5$ and $t''=0$. As typical band parameters we choose 
$t'/t=0.35$ and $t''/t=0$, which were employed to discuss 
Sr$_{3}$Ru$_{2}$O$_{7}$.\cite{yamase07b,yamase07c} 
We define the characteristic scale of $\epsilon_{\vk}^{z}$ as 
the bilayer splitting at the saddle points of the in-plane 
band dispersion, namely 
\be
\Lambda_{z}=|\epsilon_{\vk}^{z}|\; {\rm at\ the\ saddle\ points\ of} \; \epsilon_{\vk}\,.
\ee
This energy scale $\Lambda_{z}$ plays a crucial role to understand 
the property of a phase diagram as we will show below. 
The results for $\Lambda_{z}\ne 0$ are presented in Sec.III A and 
those for $\Lambda_{z}=0$ in Sec.III B. 
We deal with the case where the saddle points of $\epsilon_{\vk}$ are shifted from 
$(\pi,0)$ and $(0,\pi)$ in Sec.III~C. 
Considering the experimental situation 
in the bilayer ruthenate, we include a magnetic field in Hamiltonian (\ref{model}) and 
clarify the effect of the field in Sec.III~D. 
We set $t=1$ and measure energy in units of $t$.

\subsection{Finite $\boldsymbol{\Lambda_{z}}$} 
\begin{figure}[ht]
\centerline{\includegraphics[width=0.7\textwidth]{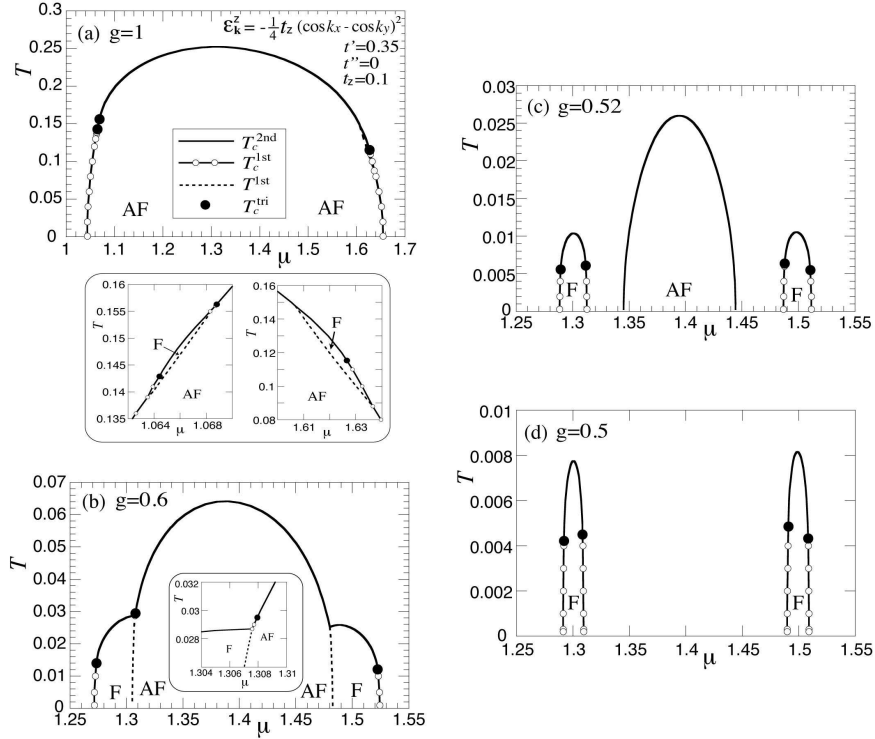}}
\caption{$\mu$-$T$ phase diagrams for the bilayer dispersion 
$\epsilon_{\vk}^{z}=-\frac{t_{z}}{4}(\cos k_x - \cos k_y)^2$ 
for several choices of $g$. 
Solid lines, $T_{c}^{\rm 2nd}$, denote second order transitions, 
while first order transitions are denoted by open circles, 
$T_{c}^{\rm 1st}$, and dotted lines, $T^{\rm 1st}$; 
the latter, present in panels (a) and (b), 
corresponds to a transition between F and AF;  
solid circles represent tricritical points. 
The insets 
magnify the regions around $\mu\approx 1.065$ and $T\approx 0.15$, 
and $\mu\approx 1.62$ and $T\approx 0.12$ in (a), 
and around $\mu\approx 1.308$ and $T\approx 0.03$ in (b).
}
\label{z1phase}
\end{figure}

We present results for the bilayer dispersion, 
$\epsilon_{\vk}^{z}=-\frac{t_{z}}{4}(\cos k_{x} - \cos k_{y})^{2}$, 
for which $\Lambda_{z}=t_z=0.1$.  
Figure~\ref{z1phase}(a) is the phase diagram for $g=1$. 
The $d$FSD instability occurs around the van Hove energy 
of the in-plane band dispersion, namely around $\mu_{\rm vH}^{0}=4t'=1.4$ 
with a dome-shaped transition line, 
as in the case of the single-layer model (Ref.~\onlinecite{yamase05}); 
$T_{c}$ is almost unchanged  by the presence of weak interlayer coupling. 
The phase diagram is almost symmetric with respect to 
$\mu=\mu_{\rm vH}^{0}$ and becomes symmetric for $t'=t''=0$ 
because of particle-hole symmetry. 
The transition is of second order at high temperature and changes to a 
first order at low temperature; the end points of the second order line 
are tricritical points. 
The AF $d$FSD is stabilized in most of the region of the phase diagram 
whereas the F $d$FSD is realized in 
very small regions near the tricritical points as shown in the inset. 
Upon decreasing the coupling constant $g$, the F region tends to be stabilized more 
near the edges of the transition line [\fig{z1phase}(b)] and eventually 
splits  from the AF region [\fig{z1phase}(c)]. 
Yet a major stacking pattern is still AF. 
Below $g=0.5$, however, the AF region disappears suddenly and no instability 
occurs around $\mu_{\rm vH}^{0}=1.4$. 
Instead the $d$FSD instability occurs 
around the van Hove energy of the bonding and antibonding bands, 
i.e. $\mu_{\rm vH}=4t'\pm t_{z}=1.3$ and $1.5$  [\fig{z1phase}(d)], 
and the phase diagram contains only the F $d$FSD.

The phase diagrams in \fig{z1phase} are strikingly similar to those for 
$\epsilon_{\vk}^{z}=-t_{z}$ (Fig.~1 in Ref.~\onlinecite{yamase09a}) regardless of 
the difference of the bilayer dispersion. 
In fact, \fig{z1phase} shows generic phase diagrams for a bilayer dispersion 
which fulfills $\Lambda_{z}\neq0$. 
Typically the AF $d$FSD state is obtained as a major stacking pattern 
when the instability occurs around the van Hove energy of 
the in-plane band dispersion. 
When the temperature scale of the $d$FSD gets smaller upon reducing $g$, 
the energy scale of the bilayer dispersion, namely $\Lambda_{z}$, becomes relevant. 
Eventually the instability occurs only around the van Hove energy of 
the bonding and antibonding bands, which is located at 
$\mu_{\rm vH}=\mu_{\rm vH}^{0}\pm\Lambda_{z}$ unless the bilayer dispersion 
shifts the saddle points of the in-plane band dispersion; 
the phase diagram is occupied only by F stacking. 
Therefore as long as $\Lambda_{z}\ne 0$,  
F stacking is stabilized  as a dominant pattern 
only when the temperature scale of the $d$FSD becomes much smaller than 
$\Lambda_{z}$, and otherwise the major stacking is AF.

\subsection{$\boldsymbol{\Lambda_{z}=0}$}
\begin{figure}[ht]
\centerline{\includegraphics[width=0.7\textwidth]{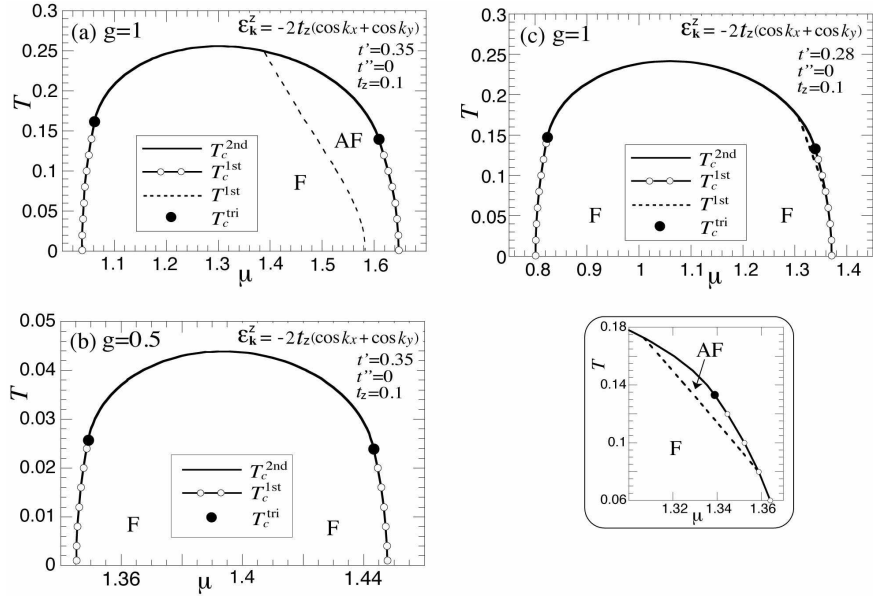}}
\caption{$\mu$-$T$ phase diagrams for the bilayer dispersion 
$\epsilon_{\vk}^{z}=-2t_{z}( \cos k_x + \cos k_y)$ for $g=1$ (a) and $0.5$ (b). 
The notation is the same as in Fig.~3. 
In (c), the hopping integral is reduced to $t'=0.28$ and thus 
the van Hove energy of the in-plane band shifts from 
$\mu_{\rm vH}^{0}=1.4$ to $\mu_{\rm vH}^{0}=1.12$. 
The inset magnifies the region near the tricritical point at higher $\mu$ in (c). 
} 
\label{z3phase}
\end{figure}

In the case of $\Lambda_{z}=0$, 
we cannot extract a generic conclusion about 
the phase diagram of the $d$FSD. The result depends strongly on the form of 
a bilayer dispersion. 
We first consider the bilayer dispersion 
$\epsilon_{\vk}^{z} = - 2t_{z} (\cos k_{x} + \cos k_{y})$, 
which is the simplest one fulfilling $\Lambda_{z}=0$. 
The obtained phase diagrams are shown in \fig{z3phase}. 
In contrast to \fig{z1phase}, 
we see that the instability occurs around $\mu_{\rm vH}^{0}=1.4$ even for a small $g$. 
This is because the bonding and antibonding bands retain the same van Hove 
energy as that of in-plane band dispersion, namely $\mu_{\rm vH}=1.4$, for 
the present bilayer dispersion. We always obtain the F $d$FSD as a major stacking pattern 
for both $g=1$ [\fig{z3phase}(a)] and $0.5$  [\fig{z3phase}(b)]. 
While one would see a sizable region of AF stacking for a large $g$ [\fig{z3phase}(a)], 
this AF region results from the presence of a large $t'$. 
In fact, AF stacking is strongly suppressed for a smaller $t'$ [\fig{z3phase}(c)] 
and completely disappears for $t' \lesssim 0.27$. 
F stacking then prevails in the whole region of the phase diagram even for $g=1$. 

\begin{figure}[ht]
\centerline{\includegraphics[width=0.35\textwidth]{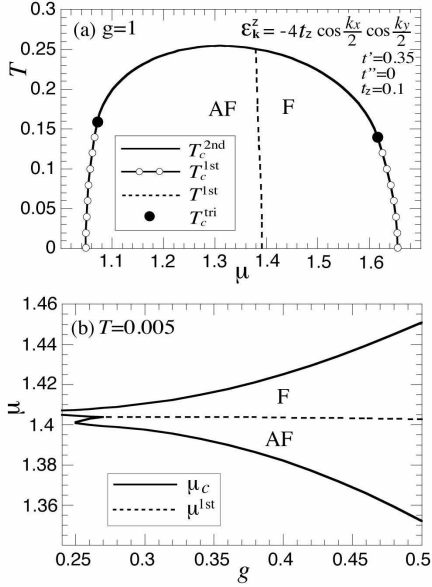}}
\caption{(a) $\mu$-$T$ phase diagrams for the bilayer dispersion 
$\epsilon_{\vk}^{z}=-4t_{z} \cos\frac{k_x}{2} \cos\frac{k_y}{2}$ for $g=1$.  
The notation is the same as in Fig.~3. 
(b) $g$ dependence of $\mu_{c}$, 
the critical chemical potential of the $d$FSD instability, at $T=0.005$. 
The transition is of first order, except for 
$(g,\mu)\approx (0.25, 1.401)$ and $(0.24, 1.406)$, 
where the second order transition occurs. 
$\mu^{\rm 1st}$ denotes a first order transition inside the 
symmetry-broken phase.
} 
\label{z2phase}
\end{figure}

The condition $\Lambda_{z}=0$ also holds for the bilayer dispersion 
$\epsilon_{\vk}^{z}= -4t_{z} \cos\frac{k_x}{2} \cos\frac{k_y}{2}$, which is   
expected for a system where lattice sites shift by $(\frac{1}{2}, \frac{1}{2})$ 
between adjacent layers. 
Figure~\ref{z2phase}(a) shows that the $d$FSD instability occurs around 
$\mu_{\rm vH}^{0}=1.4$, the same as \fig{z3phase}. 
However, the stacking pattern is qualitatively different from \fig{z3phase}. 
AF stacking is stabilized for $\mu \lesssim \mu_{\rm vH}^{0}$ while 
F stacking for $\mu \gtrsim \mu_{\rm vH}^{0}$. 
This property does not change for a smaller $g$ as shown in \fig{z2phase}(b). 
For a much smaller $g <0.25$, however, 
the region of AF stacking shrinks to disappear 
and F stacking always becomes dominant [\fig{z2phase}(b)]. 
This is because  both bonding and antibonding bands have the van Hove energy 
at  $\mu_{\rm vH}=\mu_{\rm vH}^{0}+\frac{t_{z}^{2}}{t+2t'} \approx 1.4059$.   
The shift of the van Hove energy from $\mu_{\rm vH}^{0}$ is very small and 
thus such an effect starts to appear only when the temperature scale of the $d$FSD 
is substantially reduced to become comparable to that of $|\mu_{\rm vH}-\mu_{\rm vH}^{0}|$.  
In this case, as in the case of \fig{z1phase}(d), the $d$FSD instability occurs around 
$\mu_{\rm vH}$, leading to F stacking. 
Since both bonding and antibonding bands have the same van Hove energy 
for the present bilayer dispersion, only one F stacking region is obtained in \fig{z2phase}(b).

\subsection{Saddle points away from $\boldsymbol{(\pi,0)}$ and $\boldsymbol{(0,\pi)}$}
\begin{figure}[ht]
\centerline{\includegraphics[width=0.7\textwidth]{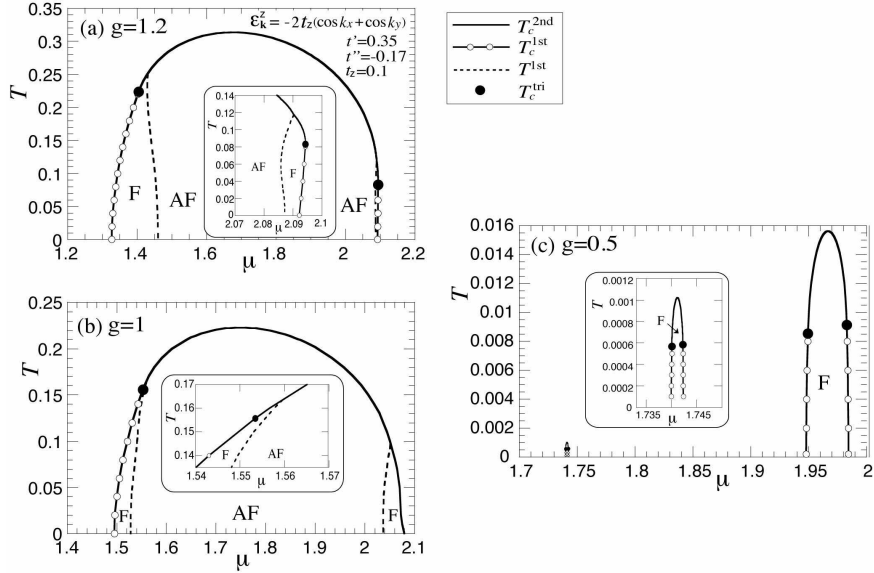}}
\caption{(a) $\mu$-$T$ phase diagrams for the bilayer dispersion 
$\epsilon_{\vk}^{z}=-2t_{z} (\cos k_x + \cos k_y)$ for $g=1.2$ (a), 
$1.0$ (b), and $0.5$ (c). 
The notation is the same as in Fig.~3; dotted lines ($T^{\rm 1st}$) are 
present only in (a) and (b). 
The inset in (a) magnifies the region near the first order transition at higher $\mu$; 
in (b) the inset clarifies the region near the tricritical point around 
$\mu=1.55$ and $T=0.156$; 
the inset in (c) magnifies the phase at lower $\mu$. 
}
\label{z3phase2}
\end{figure}

When a moderate $t''$ is introduced, the saddle points of the in-plane band dispersion shifts 
to $(0, \cos^{-1}\alpha)$ or  $(\pi, \cos^{-1}\beta)$ with 
$\alpha=-\frac{t+2t'}{4t''}$ and $\beta=-\frac{t-2t'}{4t''}$ if $|\alpha| <1$ or $|\beta|<1$. 
In this subsection, we present the results for $t'=0.35$ and $t''=-0.17$. 
The saddle points of $\epsilon_{\vk}$ are then in 
$(\pi, \pm\phi)$ and $(\pm \phi, \pi)$ with 
$\phi =\cos^{-1}\beta \approx 0.35\pi$. 
As a bilayer dispersion we employ 
$\epsilon_{\vk}^{z}=-2t_{z}(\cos k_{x} + \cos k_{y})$, for which $\Lambda_{z}$ now becomes 
finite, i.e., $\Lambda_{z}\approx 0.11$. 
Figure~\ref{z3phase2}(a) is a result for a large $g (=1.2)$. 
The presence of the finite $\Lambda_{z}$ yields the result completely different from \fig{z3phase}(a), 
although the same bilayer dispersion is employed; 
the major stacking pattern now becomes AF. 
The phase diagram in turn becomes very similar to the case of $\Lambda_{z} \ne 0$, 
i.e., \fig{z1phase}(a), regardless of 
difference of band parameters and a bilayer dispersion. 
This demonstrates the importance to recognize whether the energy scale of $\Lambda_{z}$ 
is finite or vanishes, in order to understand 
the phase diagram in the bilayer model of the $d$FSD. 

While a major phase is AF for $g=1$, the second order transition line extends down to 
$T=0$, leading to a quantum phase transition to the $d$FSD state. 
This property does not come from the bilayer effect, 
but from the additional singularity, namely the jump, of the density of state at $\mu=2.08$, 
due to the local extremes of the in-plane band dispersion at $(\pi, 0)$ and $(0, \pi)$. 
This quantum phase transition is realized as long as the $d$FSD instability 
occurs near the chemical potential corresponding to the jump of the 
density of states; in the present case, we obtain a quantum phase transition 
for $1.1 \gtrsim g \gtrsim 0.65$.\cite{miscqcp}    
Except for this, 
the phase diagram has qualitatively same properties as \fig{z1phase}(a). 

For smaller $g$, the temperature scale of the $d$FSD 
becomes small, and the other energy scale set by $\Lambda_{z}$ should be taken into account. 
Figure~\ref{z3phase2}(c) is the result for $g=0.5$. 
The $d$FSD instability occurs around the van Hove energy of the bonding 
and antibonding bands, i.e., 
$\mu_{\rm vH}=(2-\beta)(t \mp t_{z})+2t'\beta=1.742$ and $1.965$, respectively. 
In this case, as already shown in \fig{z1phase}(d), the phase diagram is occupied 
by the F stacking. 
The $d$FSD instability around $\mu_{\rm vH}=1.742$ is strongly suppressed 
compared to that around $\mu_{\rm vH}=1.965$. 
This asymmetry comes from strong breaking of particle-hole 
symmetry due to the presence of sizable $t'$ and $t''$.

\subsection{Connection to 
Sr$\boldsymbol{_{3}}$Ru$\boldsymbol{_{2}}$O$\boldsymbol{_{7}}$} 
The bilayer ruthenate Sr$_{3}$Ru$_{2}$O$_{7}$ is 
a material expected to exhibit the $d$FSD instability,\cite{grigera04,borzi07}  
which is also suggested by theoretical studies.\cite{kee05,doh07,yamase07b,yamase07c,ho08} 
Its experimental phase diagram was obtained as a function of a magnetic field. 
We thus include a magnetic field 
\be
-h \sum_{\vk, \sigma, i} \sigma c_{\vk \sigma}^{i \dagger} c_{\vk \sigma}^{i} 
\ee
in Hamiltonian (\ref{fkk}). 
Following the previous theoretical work in the single-layer model\cite{yamase07b,yamase07c}  
and LDA calculations\cite{liu08} for Sr$_{3}$Ru$_{2}$O$_{7}$, 
we choose the band parameters $t'=0.35$, $t''=0$,\cite{miscarpes}  
and $\epsilon_{\vk}^{z}=-t_{z}$ with $t_{z}=0.1$; $\Lambda_{z}$ becomes finite. 
Since the temperature scale of the $d$FSD instability in Sr$_{3}$Ru$_{2}$O$_{7}$ 
is about 1 K and is expected much smaller than $\Lambda_{z} (=0.1)$, 
we imitate such a situation choosing a small coupling constant $g=0.5$. 
We set the chemical potential to $\mu=1.288$ so that the $d$FSD 
instability occurs when a magnetic field is applied, 
modeling the experimental situation.\cite{grigera04,borzi07} 
Since the phase diagram is symmetric with respect to $h \rightarrow -h$ and 
$\sigma \rightarrow -\sigma$, we focus on the region $h>0$. 

\begin{figure}[t]
\centerline{\includegraphics[width=0.35\textwidth]{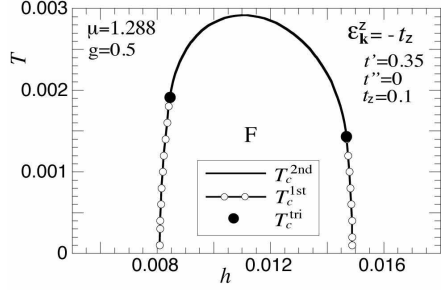}}
\caption{$h$-$T$ phase diagram designed for Sr$_{3}$Ru$_{2}$O$_{7}$; 
$\epsilon_{\vk}^{z}=-t_{z}$, $g=0.5$, and $\mu=1.288$. 
} 
\label{z0phase2}
\end{figure}

Figure~\ref{z0phase2} is the obtained phase diagram, whose property is the same as 
that obtained in the single-layer model.\cite{yamase07b,yamase07c} 
The instability occurs around the van Hove energy of the bonding band of up-spins, 
i.e., $h=0.012$. 
As shown in \fig{z1phase}(d), the phase diagram in this case is 
occupied by the F region. 
By analogy with \fig{z1phase}(d), another F $d$FSD phase is expected 
around the van Hove energy of the antibonding band of up-spins, which 
is located at $h=0.212$. 
However, as clarified in Ref.~\onlinecite{yamase07c}, 
a magnetic field strongly suppresses the onset temperature of the $d$FSD. 
In the present parameters, the maximal $T_c$ of the second $d$FSD phase 
becomes less than $0.0002$, one order magnitude smaller than \fig{z0phase2}. 
Moreover, the field range of the second $d$FSD phase is less than about 0.0002 around 
$h=0.212$.  
In experiments, therefore, the detection of the predicted second $d$FSD phase 
requires not only a measurement at a very low temperature much less than 1 K, 
but also very precise tuning of a magnetic field.

\section{Conclusions} 
We have performed a comprehensive study of the $d$FSD instability in the 
bilayer model considering various bilayer dispersions and tuning 
coupling strength and long-range hopping integrals. 
The important quantity is $\Lambda_{z}$, i.e., the energy of the bilayer splitting at 
the saddle points of the in-plane band dispersion. 
As along as $\Lambda_{z}\ne 0$, a major stacking pattern is usually AF, 
and F stacking is stabilized as a dominant pattern 
only when the temperature scale of the $d$FSD becomes much smaller 
than $\Lambda_{z}$. 
For $\Lambda_{z}=0$, the phase diagram 
depends strongly on the choice of the bilayer dispersion, leading to 
a variety of phase diagrams. 
These conclusion holds even when the saddle points of the in-plane band dispersion 
shift from $(\pi, 0)$ and $(0, \pi)$. 
In connection with Sr$_{3}$Ru$_{2}$O$_{7}$, 
the effect of a magnetic field on the bilayer model is studied. 
Since in Sr$_{3}$Ru$_{2}$O$_{7}$, 
we expect $\Lambda_{z}\ne 0$ and the temperature scale of the $d$FSD 
is likely much smaller than $\Lambda_{z}$, 
we predict the F $d$FSD instability 
around the van Hove energy of, in principle, both bonding and 
antibonding bands. 
However, the $d$FSD phase around the antibonding band 
turns out to be strongly suppressed by a magnetic field. 

In Sr$_{3}$Ru$_{2}$O$_{7}$, there are three different orbitals, 
$d_{xy}$, $d_{yz}$, and $d_{xz}$ in Ru sites, all of which 
form the bands crossing the Fermi energy. 
Previous theoretical studies\cite{kee05,doh07,yamase07b,yamase07c,ho08} 
as well as the present work 
are based on the assumption that $d_{xy}$ orbitals form an active band 
of the $d$FSD instability. 
Recently Raghu {\it et al.},\cite{raghu09} and Lee and Wu\cite{wclee09} 
proposed a different scenario that $d_{yz}$ and $d_{xz}$ orbitals are 
responsible for the $d$FSD instability.  
Considering a rich variety of phase diagrams obtained in the present bilayer model, 
it may be  worth investigating a role of weak bilayer coupling in their scenario.

Implications of the present results for cuprate superconductors 
may be obtained from the analysis of the $t$-$J$ model. 
The $t$-$J$ model contains the effective interaction described by Eqs.(1) and (2) with 
a coupling constant $g=3J/8$.\cite{yamase00b} While $g$ seems small,  the highest critical 
temperature of the $d$FSD instability reaches  around $\sim 0.2J$ 
close to half filling in the so-called uniform resonating-valence-bond state assumed 
down to zero temperature (see Sec.~3.1 and 3.3 in Ref.~\onlinecite{yamase00b}). 
This is because the nearest-neighbor hopping integral 
is strongly renormalized to become smaller than $J$ at low doping. 
Hence the $t$-$J$ model may correspond to the case of a relatively large $g (\sim 1)$ 
in the present work, implying a large effective interaction of the $d$FSD for cuprates. 

Application of the present results to cuprate superconductors, however, is not 
straightforward, because the $d$FSD is order competing with 
superconductivity as found in the $t$-$J$\cite{yamase00a,yamase00b,edegger06}  
and Hubbard\cite{metzner00,wegner02,honerkamp02,kampf03,neumayr03}  models. 
In fact,  the $d$FSD instability can be prevailed over by superconductivity. 
Nevertheless sizable correlations of the $d$FSD may survive.\cite{yamase04b}  
The $d$FSD is still an important tendency,  
leading to a giant response to a small external anisotropy. 
This idea was invoked to understand the shape of the Fermi surface and magnetic excitations 
in La-based cuprates\cite{yamase00a,yamase0107} 
as well as the strong anisotropy of magnetic excitations observed in 
YBa$_2$Cu$_3$O$_y$.\cite{yamase06,yamase09}  
Furthermore, sizable $d$FSD fluctuations substantially reduce the lifetime of quasiparticles 
in the antinodal region of the Fermi surface while not in the 
nodal direction.\cite{metzner03} In this sense, the $d$FSD fluctuations contribute 
to pseudogap behavior, which may be relevant to the strongly underdoped  YBa$_2$Cu$_3$O$_y$.\cite{yamase09}

The competition of the $d$FSD and superconductivity was studied in a general setting 
tuning coupling strength of superconductivity and turned out 
to lead to a variety of qualitatively distinct phase diagrams.\cite{yamase07a} 
Such a study may be extended to the bliayer case. 
Three energy scales of $\Lambda_{z}$, coupling strength of 
superconductivity, and that of the $d$FSD may play an important role to 
elucidate the phase diagram. 
It is also interesting  to see whether the competition with 
superconductivity favors F or AF stacking of the $d$FSD. 

Allowing a small momentum transfer in the forward scattering interaction (\ref{fkk}), 
one can incorporate fluctuations of the $d$FSD.\cite{metzner03} 
Fluctuations of the $d$FSD were studied in the context of quantum criticality,\cite{metzner03,dellanna0607}  the competition with superconductivity,\cite{yamase04b}  
and quantum phase transition deeply inside 
the $d$-wave  superconducting state.\cite{kim08,huh08} 
The present work provides a sound basis to extend such studies to a bilayer case, 
which is more realistic for various materials. 

\begin{acknowledgments}
The author is grateful to W. Metzner, 
G. Sangiovanni, and R. Zeyher for very useful discussions, 
and  to O. K. Andersen and G.-Q. Liu for extracting the tight binding parameters for 
Sr$_{3}$Ru$_{2}$O$_{7}$. 
He expresses his sincere thanks to P. Jakubczyk
for a very careful reading of the manuscript and for very thoughtful suggestions. 
\end{acknowledgments}


\bibliography{main.bib}

\begin{thebibliography}{52}
\expandafter\ifx\csname natexlab\endcsname\relax\def\natexlab#1{#1}\fi
\expandafter\ifx\csname bibnamefont\endcsname\relax
  \def\bibnamefont#1{#1}\fi
\expandafter\ifx\csname bibfnamefont\endcsname\relax
  \def\bibfnamefont#1{#1}\fi
\expandafter\ifx\csname citenamefont\endcsname\relax
  \def\citenamefont#1{#1}\fi
\expandafter\ifx\csname url\endcsname\relax
  \def\url#1{\texttt{#1}}\fi
\expandafter\ifx\csname urlprefix\endcsname\relax\def\urlprefix{URL }\fi
\providecommand{\bibinfo}[2]{#2}
\providecommand{\eprint}[2][]{\url{#2}}

\bibitem[{\citenamefont{Yamase and Kohno}(2000{\natexlab{a}})}]{yamase00a}
\bibinfo{author}{\bibfnamefont{H.}~\bibnamefont{Yamase}} \bibnamefont{and}
  \bibinfo{author}{\bibfnamefont{H.}~\bibnamefont{Kohno}},
  \bibinfo{journal}{J.\ Phys.\ Soc.\ Jpn.} \textbf{\bibinfo{volume}{69}},
  \bibinfo{pages}{332} (\bibinfo{year}{2000}{\natexlab{a}}).

\bibitem[{\citenamefont{Yamase and Kohno}(2000{\natexlab{b}})}]{yamase00b}
\bibinfo{author}{\bibfnamefont{H.}~\bibnamefont{Yamase}} \bibnamefont{and}
  \bibinfo{author}{\bibfnamefont{H.}~\bibnamefont{Kohno}},
  \bibinfo{journal}{J.\ Phys.\ Soc.\ Jpn.} \textbf{\bibinfo{volume}{69}},
  \bibinfo{pages}{2151} (\bibinfo{year}{2000}{\natexlab{b}}).

\bibitem[{\citenamefont{Halboth and Metzner}(2000)}]{metzner00}
\bibinfo{author}{\bibfnamefont{C.~J.} \bibnamefont{Halboth}} \bibnamefont{and}
  \bibinfo{author}{\bibfnamefont{W.}~\bibnamefont{Metzner}},
  \bibinfo{journal}{Phys.\ Rev.\ Lett.} \textbf{\bibinfo{volume}{85}},
  \bibinfo{pages}{5162} (\bibinfo{year}{2000}).

\bibitem[{\citenamefont{Valenzuela and Vozmediano}(2001)}]{valenzuela01}
\bibinfo{author}{\bibfnamefont{B.}~\bibnamefont{Valenzuela}} \bibnamefont{and}
  \bibinfo{author}{\bibfnamefont{M.~A.~H.} \bibnamefont{Vozmediano}},
  \bibinfo{journal}{Phys.\ Rev.\ B} \textbf{\bibinfo{volume}{63}},
  \bibinfo{pages}{153103} (\bibinfo{year}{2001}).

\bibitem[{\citenamefont{Pomeranchuk}(1958)}]{pomeranchuk58}
\bibinfo{author}{\bibfnamefont{I.~J.} \bibnamefont{Pomeranchuk}},
  \bibinfo{journal}{Sov.\ Phys.\ JETP} \textbf{\bibinfo{volume}{8}},
  \bibinfo{pages}{361} (\bibinfo{year}{1958}).

\bibitem[{\citenamefont{Miyanaga and Yamase}(2006)}]{miyanaga06}
\bibinfo{author}{\bibfnamefont{A.}~\bibnamefont{Miyanaga}} \bibnamefont{and}
  \bibinfo{author}{\bibfnamefont{H.}~\bibnamefont{Yamase}},
  \bibinfo{journal}{Phys. Rev. B} \textbf{\bibinfo{volume}{73}},
  \bibinfo{pages}{174513} (\bibinfo{year}{2006}).

\bibitem[{\citenamefont{{B. Edegger, V. N. Muthukumar, and C.
  Gros}}(2006)}]{edegger06}
\bibinfo{author}{\bibnamefont{{B. Edegger, V. N. Muthukumar, and C. Gros}}},
  \bibinfo{journal}{Phys.\ Rev.\ B} \textbf{\bibinfo{volume}{74}},
  \bibinfo{pages}{165109} (\bibinfo{year}{2006}).

\bibitem[{kee()}]{kee0304}
\bibinfo{note}{H.-Y. Kee, E. H. Kim, and C.-H. Chung, Phys.\ Rev.\ B {\bf 68},
  245109 (2003); I. Khavkine, C.-H. Chung, V. Oganesyan, and H.-Y. Kee, {\it
  ibid.} {\bf 70}, 155110 (2004).}

\bibitem[{\citenamefont{Yamase et~al.}(2005)\citenamefont{Yamase, Oganesyan,
  and Metzner}}]{yamase05}
\bibinfo{author}{\bibfnamefont{H.}~\bibnamefont{Yamase}},
  \bibinfo{author}{\bibfnamefont{V.}~\bibnamefont{Oganesyan}},
  \bibnamefont{and} \bibinfo{author}{\bibfnamefont{W.}~\bibnamefont{Metzner}},
  \bibinfo{journal}{Phys.\ Rev.\ B} \textbf{\bibinfo{volume}{72}},
  \bibinfo{pages}{35114} (\bibinfo{year}{2005}).

\bibitem[{\citenamefont{Kivelson et~al.}(1998)\citenamefont{Kivelson, Fradkin,
  and Emery}}]{kivelson98}
\bibinfo{author}{\bibfnamefont{S.~A.} \bibnamefont{Kivelson}},
  \bibinfo{author}{\bibfnamefont{E.}~\bibnamefont{Fradkin}}, \bibnamefont{and}
  \bibinfo{author}{\bibfnamefont{V.~J.} \bibnamefont{Emery}},
  \bibinfo{journal}{Nature (London)} \textbf{\bibinfo{volume}{393}},
  \bibinfo{pages}{550} (\bibinfo{year}{1998}).

\bibitem[{\citenamefont{Kivelson et~al.}(2003)\citenamefont{Kivelson, Bindloss,
  Fradkin, Oganesyan, Tranquada, Kapitulnik, and Howald}}]{kivelson03}
\bibinfo{author}{\bibfnamefont{S.~A.} \bibnamefont{Kivelson}},
  \bibinfo{author}{\bibfnamefont{I.~P.} \bibnamefont{Bindloss}},
  \bibinfo{author}{\bibfnamefont{E.}~\bibnamefont{Fradkin}},
  \bibinfo{author}{\bibfnamefont{V.}~\bibnamefont{Oganesyan}},
  \bibinfo{author}{\bibfnamefont{J.~M.} \bibnamefont{Tranquada}},
  \bibinfo{author}{\bibfnamefont{A.}~\bibnamefont{Kapitulnik}},
  \bibnamefont{and} \bibinfo{author}{\bibfnamefont{C.}~\bibnamefont{Howald}},
  \bibinfo{journal}{Rev. \ Mod.\ Phys.} \textbf{\bibinfo{volume}{75}},
  \bibinfo{pages}{1201} (\bibinfo{year}{2003}).

\bibitem[{\citenamefont{{C. A. Lamas, D. C. Cabra, and N.
  Grandi}}(2008)}]{lamas08}
\bibinfo{author}{\bibnamefont{{C. A. Lamas, D. C. Cabra, and N. Grandi}}},
  \bibinfo{journal}{Phys.\ Rev.\ B} \textbf{\bibinfo{volume}{78}},
  \bibinfo{pages}{115104} (\bibinfo{year}{2008}).

\bibitem[{\citenamefont{{S. A. Grigera, P. Gegenwart, R. A. Borzi, F. Weickert,
  A. J. Schofield, R. S. Perry, T. Tayama, T. Sakakibara, Y. Maeno, A. G.
  Green, and A. P. Mackenzie}}(2004)}]{grigera04}
\bibinfo{author}{\bibnamefont{{S. A. Grigera, P. Gegenwart, R. A. Borzi, F.
  Weickert, A. J. Schofield, R. S. Perry, T. Tayama, T. Sakakibara, Y. Maeno,
  A. G. Green, and A. P. Mackenzie}}}, \bibinfo{journal}{Science}
  \textbf{\bibinfo{volume}{306}}, \bibinfo{pages}{1154} (\bibinfo{year}{2004}).

\bibitem[{\citenamefont{{R. A. Borzi, S. A. Grigera, J. Farrell, R. S. Perry,
  S. J. S. Lister, S. L. Lee, D. A. Tennant, Y. Maeno, and A. P.
  Mackenzie}}(2007)}]{borzi07}
\bibinfo{author}{\bibnamefont{{R. A. Borzi, S. A. Grigera, J. Farrell, R. S.
  Perry, S. J. S. Lister, S. L. Lee, D. A. Tennant, Y. Maeno, and A. P.
  Mackenzie}}}, \bibinfo{journal}{Science} \textbf{\bibinfo{volume}{315}},
  \bibinfo{pages}{214} (\bibinfo{year}{2007}).

\bibitem[{\citenamefont{Kee and Kim}(2005)}]{kee05}
\bibinfo{author}{\bibfnamefont{H.-Y.} \bibnamefont{Kee}} \bibnamefont{and}
  \bibinfo{author}{\bibfnamefont{Y.~B.} \bibnamefont{Kim}},
  \bibinfo{journal}{Phys.\ Rev.\ B} \textbf{\bibinfo{volume}{71}},
  \bibinfo{pages}{184402} (\bibinfo{year}{2005}).

\bibitem[{\citenamefont{Doh et~al.}(2007)\citenamefont{Doh, Kim, and
  Ahn}}]{doh07}
\bibinfo{author}{\bibfnamefont{H.}~\bibnamefont{Doh}},
  \bibinfo{author}{\bibfnamefont{Y.~B.} \bibnamefont{Kim}}, \bibnamefont{and}
  \bibinfo{author}{\bibfnamefont{K.~H.} \bibnamefont{Ahn}},
  \bibinfo{journal}{Phys.\ Rev.\ Lett.} \textbf{\bibinfo{volume}{98}},
  \bibinfo{pages}{126407} (\bibinfo{year}{2007}).

\bibitem[{\citenamefont{Yamase and Katanin}(2007)}]{yamase07b}
\bibinfo{author}{\bibfnamefont{H.}~\bibnamefont{Yamase}} \bibnamefont{and}
  \bibinfo{author}{\bibfnamefont{A.~A.} \bibnamefont{Katanin}},
  \bibinfo{journal}{J.\ Phys.\ Soc.\ Jpn.} \textbf{\bibinfo{volume}{76}},
  \bibinfo{pages}{073706} (\bibinfo{year}{2007}).

\bibitem[{\citenamefont{Yamase}(2007)}]{yamase07c}
\bibinfo{author}{\bibfnamefont{H.}~\bibnamefont{Yamase}},
  \bibinfo{journal}{Phys.\ Rev.\ B} \textbf{\bibinfo{volume}{76}},
  \bibinfo{pages}{155117} (\bibinfo{year}{2007}).

\bibitem[{\citenamefont{Ho and Schofield}(2008)}]{ho08}
\bibinfo{author}{\bibfnamefont{A.~F.} \bibnamefont{Ho}} \bibnamefont{and}
  \bibinfo{author}{\bibfnamefont{A.~J.} \bibnamefont{Schofield}},
  \bibinfo{journal}{Europhys. Lett.} \textbf{\bibinfo{volume}{84}},
  \bibinfo{pages}{27007} (\bibinfo{year}{2008}).

\bibitem[{\citenamefont{Yamase and Metzner}(2006)}]{yamase06}
\bibinfo{author}{\bibfnamefont{H.}~\bibnamefont{Yamase}} \bibnamefont{and}
  \bibinfo{author}{\bibfnamefont{W.}~\bibnamefont{Metzner}},
  \bibinfo{journal}{Phys.\ Rev.\ B} \textbf{\bibinfo{volume}{73}},
  \bibinfo{pages}{214517} (\bibinfo{year}{2006}).

\bibitem[{\citenamefont{Hinkov et~al.}(2004)\citenamefont{Hinkov, Pailh\`{e}s,
  Bourges, Sidis, Ivanov, Kulakov, Lin, Chen, Bernhard, and Keimer}}]{hinkov04}
\bibinfo{author}{\bibfnamefont{V.}~\bibnamefont{Hinkov}},
  \bibinfo{author}{\bibfnamefont{S.}~\bibnamefont{Pailh\`{e}s}},
  \bibinfo{author}{\bibfnamefont{P.}~\bibnamefont{Bourges}},
  \bibinfo{author}{\bibfnamefont{Y.}~\bibnamefont{Sidis}},
  \bibinfo{author}{\bibfnamefont{A.}~\bibnamefont{Ivanov}},
  \bibinfo{author}{\bibfnamefont{A.}~\bibnamefont{Kulakov}},
  \bibinfo{author}{\bibfnamefont{C.~T.} \bibnamefont{Lin}},
  \bibinfo{author}{\bibfnamefont{D.}~\bibnamefont{Chen}},
  \bibinfo{author}{\bibfnamefont{C.}~\bibnamefont{Bernhard}}, \bibnamefont{and}
  \bibinfo{author}{\bibfnamefont{B.}~\bibnamefont{Keimer}},
  \bibinfo{journal}{Nature (London)} \textbf{\bibinfo{volume}{430}},
  \bibinfo{pages}{650} (\bibinfo{year}{2004}).

\bibitem[{\citenamefont{{V. Hinkov, P. Bourges, S. Pailh\`{e}s, Y. Sidis, A.
  Ivanov, C. D. Frost, T. G. Perring, C. T. Lin, D. P. Chen, and B.
  Keimer}}(2007)}]{hinkov07}
\bibinfo{author}{\bibnamefont{{V. Hinkov, P. Bourges, S. Pailh\`{e}s, Y. Sidis,
  A. Ivanov, C. D. Frost, T. G. Perring, C. T. Lin, D. P. Chen, and B.
  Keimer}}}, \bibinfo{journal}{Nat. Phys.} \textbf{\bibinfo{volume}{3}},
  \bibinfo{pages}{780} (\bibinfo{year}{2007}).

\bibitem[{\citenamefont{{V. Hinkov, D. Haug, B. Fauqu\'{e}, P. Bourges, Y.
  Sidis, A. Ivanov, C. Bernhard, C. T. Lin, and B. Keimer}}(2008)}]{hinkov08}
\bibinfo{author}{\bibnamefont{{V. Hinkov, D. Haug, B. Fauqu\'{e}, P. Bourges,
  Y. Sidis, A. Ivanov, C. Bernhard, C. T. Lin, and B. Keimer}}},
  \bibinfo{journal}{Science} \textbf{\bibinfo{volume}{319}},
  \bibinfo{pages}{597} (\bibinfo{year}{2008}).

\bibitem[{\citenamefont{{E.-A. Kim, M. J. Lawler, P. Oreto, S. Sachdev, E.
  Fradkin, and S. A. Kivelson}}(2008)}]{kim08}
\bibinfo{author}{\bibnamefont{{E.-A. Kim, M. J. Lawler, P. Oreto, S. Sachdev,
  E. Fradkin, and S. A. Kivelson}}}, \bibinfo{journal}{Phys.\ Rev.\ B}
  \textbf{\bibinfo{volume}{77}}, \bibinfo{pages}{184514}
  (\bibinfo{year}{2008}).

\bibitem[{\citenamefont{{Y. Huh and S. Sachdev}}(2008)}]{huh08}
\bibinfo{author}{\bibnamefont{{Y. Huh and S. Sachdev}}},
  \bibinfo{journal}{Phys.\ Rev.\ B} \textbf{\bibinfo{volume}{78}},
  \bibinfo{pages}{064512} (\bibinfo{year}{2008}).

\bibitem[{\citenamefont{Yamase}(2009{\natexlab{a}})}]{yamase09}
\bibinfo{author}{\bibfnamefont{H.}~\bibnamefont{Yamase}},
  \bibinfo{journal}{Phys.\ Rev.\ B} \textbf{\bibinfo{volume}{79}},
  \bibinfo{pages}{052501} (\bibinfo{year}{2009}{\natexlab{a}}).

\bibitem[{\citenamefont{{M. P. Lilly, K. B. Cooper, J. P. Eisenstein, L. N.
  Pfeiffer, and K. W. West}}(1999{\natexlab{a}})}]{lilly99}
\bibinfo{author}{\bibnamefont{{M. P. Lilly, K. B. Cooper, J. P. Eisenstein, L.
  N. Pfeiffer, and K. W. West}}}, \bibinfo{journal}{Phys.\ Rev.\ Lett.}
  \textbf{\bibinfo{volume}{82}}, \bibinfo{pages}{394}
  (\bibinfo{year}{1999}{\natexlab{a}}).

\bibitem[{\citenamefont{{R. R. Du, D. C. Tsui, H. L. Stormer, L. N. Pfeiffer,
  K. W. Baldwin, K. W. West}}(1999)}]{du99}
\bibinfo{author}{\bibnamefont{{R. R. Du, D. C. Tsui, H. L. Stormer, L. N.
  Pfeiffer, K. W. Baldwin, K. W. West}}}, \bibinfo{journal}{Solid State
  Commun.} \textbf{\bibinfo{volume}{109}}, \bibinfo{pages}{389}
  (\bibinfo{year}{1999}).

\bibitem[{\citenamefont{{M. P. Lilly, K. B. Cooper, J. P. Eisenstein, L. N.
  Pfeiffer, and K. W. West}}(1999{\natexlab{b}})}]{lilly99a}
\bibinfo{author}{\bibnamefont{{M. P. Lilly, K. B. Cooper, J. P. Eisenstein, L.
  N. Pfeiffer, and K. W. West}}}, \bibinfo{journal}{Phys.\ Rev.\ Lett.}
  \textbf{\bibinfo{volume}{83}}, \bibinfo{pages}{824}
  (\bibinfo{year}{1999}{\natexlab{b}}).

\bibitem[{\citenamefont{{E. Fradkin and S. A. Kivelson}}(1999)}]{fradkin99}
\bibinfo{author}{\bibnamefont{{E. Fradkin and S. A. Kivelson}}},
  \bibinfo{journal}{Phys.\ Rev.\ B} \textbf{\bibinfo{volume}{59}},
  \bibinfo{pages}{8065} (\bibinfo{year}{1999}).

\bibitem[{\citenamefont{{E. Fradkin, S. A. Kivelson, E. Manousakis, and K.
  Nho}}(2000)}]{fradkin00}
\bibinfo{author}{\bibnamefont{{E. Fradkin, S. A. Kivelson, E. Manousakis, and
  K. Nho}}}, \bibinfo{journal}{Phys.\ Rev.\ Lett.}
  \textbf{\bibinfo{volume}{84}}, \bibinfo{pages}{1982} (\bibinfo{year}{2000}).

\bibitem[{\citenamefont{Wexler and Dorsey}(2001)}]{wexler01}
\bibinfo{author}{\bibfnamefont{C.}~\bibnamefont{Wexler}} \bibnamefont{and}
  \bibinfo{author}{\bibfnamefont{A.~T.} \bibnamefont{Dorsey}},
  \bibinfo{journal}{Phys.\ Rev.\ B} \textbf{\bibinfo{volume}{64}},
  \bibinfo{pages}{115312} (\bibinfo{year}{2001}).

\bibitem[{\citenamefont{{J. Quintanilla, M. Haque, and A. J.
  Schofield}}(2008)}]{quintanilla08}
\bibinfo{author}{\bibnamefont{{J. Quintanilla, M. Haque, and A. J.
  Schofield}}}, \bibinfo{journal}{Phys.\ Rev.\ B}
  \textbf{\bibinfo{volume}{78}}, \bibinfo{pages}{035131}
  (\bibinfo{year}{2008}).

\bibitem[{\citenamefont{Quintanilla and Schofield}(2006)}]{quintanilla06}
\bibinfo{author}{\bibfnamefont{J.}~\bibnamefont{Quintanilla}} \bibnamefont{and}
  \bibinfo{author}{\bibfnamefont{A.~J.} \bibnamefont{Schofield}},
  \bibinfo{journal}{Phys.\ Rev.\ B} \textbf{\bibinfo{volume}{74}},
  \bibinfo{pages}{115126} (\bibinfo{year}{2006}).

\bibitem[{\citenamefont{Yamase}(2009{\natexlab{b}})}]{yamase09a}
\bibinfo{author}{\bibfnamefont{H.}~\bibnamefont{Yamase}},
  \bibinfo{journal}{Phys.\ Rev.\ Lett.} \textbf{\bibinfo{volume}{102}},
  \bibinfo{pages}{116404} (\bibinfo{year}{2009}{\natexlab{b}}).

\bibitem[{\citenamefont{Andersen et~al.}(1995)\citenamefont{Andersen,
  Lichtenstein, Jepsen, and Paulsen}}]{andersen95}
\bibinfo{author}{\bibfnamefont{O.~K.} \bibnamefont{Andersen}},
  \bibinfo{author}{\bibfnamefont{A.~I.} \bibnamefont{Lichtenstein}},
  \bibinfo{author}{\bibfnamefont{O.}~\bibnamefont{Jepsen}}, \bibnamefont{and}
  \bibinfo{author}{\bibfnamefont{F.}~\bibnamefont{Paulsen}},
  \bibinfo{journal}{J.\ Phys.\ Chem.\ Solids} \textbf{\bibinfo{volume}{56}},
  \bibinfo{pages}{1573} (\bibinfo{year}{1995}).

\bibitem[{mis({\natexlab{a}})}]{miscchargeimbalance}
\bibinfo{note}{We searched for a charge imbalance solution for $t'=0.35$,
  $t''=0$, $g=1$, and $\epsilon_{\vk}^{z}=-t_{z}=-0.1$, and found it to be
  stabilized in very small regions around the first order transitions.}

\bibitem[{mis({\natexlab{b}})}]{miscqcp}
\bibinfo{note}{In the single-layer model for $t'=0.35$ and $t''=-0.17$, a
  quantum phase transition of the $d$FSD instability occurs for $0.7 \lesssim g
  \lesssim 1.15$. On the other hand, for the parameters $t'=-1/6$ and $t''=1/5$
  employed in Ref.~\onlinecite{yamase05}, the saddle points also shift from
  $(\pi, 0)$ and $(0, \pi)$, which then become local extremes. A quantum phase
  transition is found to be realized for $0.3 \lesssim g \lesssim 0.4$, which
  was overlooked in Ref.~\onlinecite{yamase05}.}

\bibitem[{liu()}]{liu08}
\bibinfo{note}{G.-Q. Liu and O. K. Andersen (private communication).}

\bibitem[{mis({\natexlab{c}})}]{miscarpes}
\bibinfo{note}{According to the recent angle-resolved photoemission
  spectroscopy for Sr$_{3}$Ru$_{2}$O$_{7}$,\cite{tamai08} the saddle points
  shift from $(\pi, 0)$ and $(0, \pi)$, implying the presence of a sizable
  $t''$. However, as we argued in Sec.III~C, our conclusion holds as long as
  $\Lambda_{z}\ne 0$.}

\bibitem[{\citenamefont{{S. Raghu, A. Paramekanti, E-.A. Kim, R. A. Borzi, S.
  Grigera, A. P. Mackenzie, and S. A. Kivelson}}(2009)}]{raghu09}
\bibinfo{author}{\bibnamefont{{S. Raghu, A. Paramekanti, E-.A. Kim, R. A.
  Borzi, S. Grigera, A. P. Mackenzie, and S. A. Kivelson}}},
  \bibinfo{journal}{Phys.\ Rev.\ B} \textbf{\bibinfo{volume}{79}},
  \bibinfo{pages}{214402} (\bibinfo{year}{2009}).

\bibitem[{wcl()}]{wclee09}
\bibinfo{note}{W.-C. Lee and C. Wu, arXiv: 0902.1337.}

\bibitem[{weg()}]{wegner02}
\bibinfo{note}{I. Grote, E. K{\"{o}}rding, and F. Wegner, J.\ Low\ Temp.\
  Phys.\ {\bf 126}, 1385 (2002); V. Hankevych, I. Grote, and F. Wegner, Phys.\
  Rev.\ B \ {\bf 66}, 094516 (2002)}.

\bibitem[{\citenamefont{Honerkamp et~al.}(2002)\citenamefont{Honerkamp,
  Salmhofer, and Rice}}]{honerkamp02}
\bibinfo{author}{\bibfnamefont{C.}~\bibnamefont{Honerkamp}},
  \bibinfo{author}{\bibfnamefont{M.}~\bibnamefont{Salmhofer}},
  \bibnamefont{and} \bibinfo{author}{\bibfnamefont{T.~M.} \bibnamefont{Rice}},
  \bibinfo{journal}{Eur. Phys. J. B} \textbf{\bibinfo{volume}{27}},
  \bibinfo{pages}{127} (\bibinfo{year}{2002}).

\bibitem[{\citenamefont{Kampf and Katanin}(2003)}]{kampf03}
\bibinfo{author}{\bibfnamefont{A.~P.} \bibnamefont{Kampf}} \bibnamefont{and}
  \bibinfo{author}{\bibfnamefont{A.~A.} \bibnamefont{Katanin}},
  \bibinfo{journal}{Phys.\ Rev.\ B} \textbf{\bibinfo{volume}{67}},
  \bibinfo{pages}{125104} (\bibinfo{year}{2003}).

\bibitem[{\citenamefont{Neumayr and Metzner}(2003)}]{neumayr03}
\bibinfo{author}{\bibfnamefont{A.}~\bibnamefont{Neumayr}} \bibnamefont{and}
  \bibinfo{author}{\bibfnamefont{W.}~\bibnamefont{Metzner}},
  \bibinfo{journal}{Phys.\ Rev.\ B} \textbf{\bibinfo{volume}{67}},
  \bibinfo{pages}{035112} (\bibinfo{year}{2003}).

\bibitem[{\citenamefont{Yamase}(2004)}]{yamase04b}
\bibinfo{author}{\bibfnamefont{H.}~\bibnamefont{Yamase}},
  \bibinfo{journal}{Phys.\ Rev.\ Lett.} \textbf{\bibinfo{volume}{93}},
  \bibinfo{pages}{266404} (\bibinfo{year}{2004}).

\bibitem[{yam()}]{yamase0107}
\bibinfo{note}{H. Yamase and H. Kohno, J.\ Phys.\ Soc.\ Jpn.\ {\bf 70}, 2733
  (2001); H. Yamase, Phys.\ Rev.\ B \ {\bf 75}, 014514 (2007).}

\bibitem[{\citenamefont{Metzner et~al.}(2003)\citenamefont{Metzner, Rohe, and
  Andergassen}}]{metzner03}
\bibinfo{author}{\bibfnamefont{W.}~\bibnamefont{Metzner}},
  \bibinfo{author}{\bibfnamefont{D.}~\bibnamefont{Rohe}}, \bibnamefont{and}
  \bibinfo{author}{\bibfnamefont{S.}~\bibnamefont{Andergassen}},
  \bibinfo{journal}{Phys.\ Rev.\ Lett.} \textbf{\bibinfo{volume}{91}},
  \bibinfo{pages}{066402} (\bibinfo{year}{2003}).

\bibitem[{\citenamefont{Yamase and Metzner}(2007)}]{yamase07a}
\bibinfo{author}{\bibfnamefont{H.}~\bibnamefont{Yamase}} \bibnamefont{and}
  \bibinfo{author}{\bibfnamefont{W.}~\bibnamefont{Metzner}},
  \bibinfo{journal}{Phys.\ Rev.\ B} \textbf{\bibinfo{volume}{75}},
  \bibinfo{pages}{155117} (\bibinfo{year}{2007}).

\bibitem[{del()}]{dellanna0607}
\bibinfo{note}{L. Dell'Anna and W. Metzner, Phys. Rev. B {\bf 73}, 045127
  (2006); Phys. Rev. Lett. {\bf 98}, 136402 (2007).}

\bibitem[{\citenamefont{{A. Tamai, M.P. Allan, J. F. Mercure, W. Meevasana, R.
  Dunkel, D. H. Lu, R. S. Perry, A. P. Mackenzie, D. J. Singh, Z.-X. Shen, and
  F. Baumberger}}(2008)}]{tamai08}
\bibinfo{author}{\bibnamefont{{A. Tamai, M.P. Allan, J. F. Mercure, W.
  Meevasana, R. Dunkel, D. H. Lu, R. S. Perry, A. P. Mackenzie, D. J. Singh,
  Z.-X. Shen, and F. Baumberger}}}, \bibinfo{journal}{Phys.\ Rev.\ Lett.}
  \textbf{\bibinfo{volume}{101}}, \bibinfo{pages}{026407}
  (\bibinfo{year}{2008}).

\end{thebibliography}

\end{document}